\documentclass[preprint2]{aastex}

\usepackage{amsmath}

\shorttitle{E1 transitions in helium-like ions}
\shortauthors{Johnson et al.}
\begin{document}

\title{E1 transitions between states with $n$=1 to 6 in helium-like
carbon, nitrogen, oxygen, neon, silicon, and argon}
\author{W. R. Johnson, I. M. Savukov, and  U. I. Safronova,}
\email{johnson@nd.edu}
\affil{
Department of Physics, 225 Nieuwland Science Hall\\ University of
Notre Dame, Notre Dame, IN 46556}
\author{A. Dalgarno}
\email{adalgarno@cfa.harvard.edu}
\affil{Harvard-Smithsonian Center for Astrophysics\\
60 Garden Street, Cambridge MA 02138}
\date{\today}
\begin{abstract}
Wavelengths and transition rates are given for E1 transitions
between singlet $^1\! S$, $^1\! P$, $^1\! D$, and $^1\! F$ states,
between triplet
$^3\! S$, $^3\! P$, and $^3\! D$ states, and between triplet $^3\! P_1$
and singlet
$^1\! S_0$ states in ions of astrophysical interest: helium-like
carbon, nitrogen, oxygen, neon, silicon, and argon. All possible
E1 transitions between states with $J\leq 3$ and $n\leq6$ are considered.
Energy levels and wave functions used in calculations of the
transition rates are obtained from relativistic
configuration-interaction calculations that include both
Coulomb and Breit interactions.
\end{abstract}

\maketitle

\section{Introduction}

     The emission lines resulting from electron capture by
multicharged ions colliding with neutral gases provide a powerful
diagnostic probe of astrophysical plasmas. X-rays seen in auroras on Jupiter
\citep{ME:83,WA:94,CR:95,KLD:98}, in comets \citep{LI:96,DET:97,LI:01}
and the X-ray background \citep{CR:00,CR:01} have been attributed
to electron capture by heavy ions.

Several laboratory studies of the X-ray emissions have been carried out
\citep{GR:00,BE:00,Lu:01,GR:01,BE:01,MA:01,FL:01,HA:01}.
The interpretation of the laboratory
data and the astrophysical data requires a reliable description
of the radiative cascade that follows the capture into excited states.

The cascade lines appearing at extreme ultraviolet wavelengths provide
additional probes of the environments in which the multicharged
heavy ions are present. For the helium-like ions interesting
differences may occur as the nuclear charge of the ions changes.
It has long been known that for ions beyond N$^{5+}$ the spin-forbidden
$2\, ^3\!P_1 - 1\, ^1\! S_0$ transition is more probable than the allowed
$2\, ^3\!P_1 - 2\, ^3\! S_1$ transition \citep{DD:69}.

In the most recent National Institute of Standards and Technology (NIST)
data compilation \citep{WFD:96},
recommended transition rates for helium-like C, N, and O are
based primarily on nonrelativistic calculations by
\citet{CT:92}, who used explicitly correlated
wave functions for singlet and triplet $S$, $P$,
and $D$ states with $n\leq 6$
to calculate energies and dipole oscillator strengths  for
$S-P$ and $P-D$ transitions in helium-like ions with $Z$ from 2 to 10.

In this paper, we extend the results of previous calculations
using relativistic configuration-interaction (CI) wave functions
and energies for
$(1snl)\, ^{2S+1}\! L_J$ states of helium-like ions with $J\leq 3$,
$n \leq 6$, and $Z=6,\, 7,\, 8,\, 10,\, 14$, and 18.
We present results here for the allowed singlet-singlet and triplet-triplet
transitions. The rates for triplet-triplet transitions
are determined by averaging the calculated rates over fine-structure
substates. Additionally, we present data for intercombination transitions
between $^3P_1$ and $^1S_0$ states.

Relativistic CI calculations of wavelengths and transition
probabilities have been carried out previously for E1, M1,
and M2 transitions between $n = 1$ and 2 states of helium-like ions
\citep{PJS:95}.
Nonrelativistic many-body perturbation theory, treating relativistic
corrections perturbatively, has also been used to
calculate energies of $(1snl)$
states with $n=2$--5 and $l=0$--2 \citep{VS:85}.
The present calculations extend previous relativistic calculations
to $n>2$ and extend previous nonrelativistic calculations to
higher values of $Z$.

The importance of accurate atomic characteristics for
astrophysics was further illustrated by
\citet{KD:01}, where the variability of cometary
X-ray emission induced by solar wind ions was studied.

\section{Calculation}
Relativistic CI calculations were introduced
to evaluate precise values of energies of $n=1$ and 2 states of helium-like ions
\citep{CCJ:93,CC:94} and used subsequently
to evaluate transition energies and transition rates between states of helium-like
ions with $n$ = 1 and 2 \citep{PJS:95}. In the present paper, we apply the
methods developed in these earlier papers to evaluate  wavelengths and rates
for E1 transitions between $(1snl)$ states
with $n\leq 6$ in various helium-like ions of astrophysical interest.
The computational method is summarized in the following paragraphs.

\subsection{\label{seca} Wave Functions and Energies}
The wave function describing a state with angular
momentum $J, M$ in a two-electron ion may be written  as
\begin{equation}
\Psi_{JM} = \sum_{i \le j} c_{ij} \Phi_{ij}(JM) ,  \label{eq1}
\end{equation}
where the quantities $c_{ij}$ are expansion coefficients and
where  $\Phi_{ij}(JM)$, the configuration state vectors, are given by
\begin{equation}
\Phi_{ij}(JM) = \eta_{ij} \sum_{m_i m_j} \langle j_i m_i , j_j m_j |
JM\rangle \; a^{\dagger}_i a^{\dagger}_j |0\rangle ,
\label{eq2}
\end{equation}
in second quantization, with
\begin{equation}
\eta_{ij} = \left\{
\begin{array}{ll}
1,  & i\ne j ,\\
1/{\sqrt 2}, & i = j .
\end{array}
\right.
  \label{eq3}
\end{equation}
In the above equations, we use subscripts $i$ to designate
quantum numbers $(n_i,j_i,l_i,m_i)$ of one-electron states.
The quantities $c_{ij}$, $\Phi_{ij}(JM)$, and $\eta_{ij}$ are
independent of magnetic quantum numbers $m_i$ and $m_j$.
To construct a state of even or odd parity, one requires the sum
of orbital angular momenta
$l_i + l_j$  to be either even or odd, respectively. From the
symmetry properties of the Clebsch-Gordan coefficients, it can be
shown that
\begin{equation}
\Phi_{ij}(JM) = (-1)^{j_i + j_j + J + 1} \Phi_{ji}(JM) . \label{eq4}
\end{equation}
This relation, in turn,  implies that $\Phi_{ii}(JM)$ vanishes unless
$J$ is even. The wave-function normalization condition has the
form
\begin{equation}
\langle\Psi_{JM}|\Psi_{JM}\rangle = \sum_{i \le j} c_{ij}^2  = 1 .
\label{eq5}
\end{equation}

Substituting $\Psi_{JM}$ into the Schr\"{o}dinger equation $(H_0
+V) \Psi_{JM} = E \Psi_{JM}$, one obtains the following set of
linear equations for the expansion coefficients $c_{ij}$:
\begin{equation}
  (\epsilon_i + \epsilon_j) c_{ij} + \sum_{k \le l}
 \eta_{ij} V_J(ij;kl)\, \eta_{kl} \, c_{kl} = E c_{ij} \,
  . \label{eq6}
\end{equation}
The potential matrix in Eq.~(\ref{eq6}) is
\begin{multline}
V_J(ij;kl) = \\
\sum_L (-1)^{j_j + j_k + L + J}
  \left\lbrace
  \begin{array}{ccc}
  j_i&j_j&J\\
  j_l&j_k&L
 \end{array}
  \right\rbrace
  X_L(ijkl)  \\
+ \sum_L (-1)^{j_j + j_k + L}
  \left\lbrace
  \begin{array}{ccc}
  j_i&j_j&J\\
  j_k&j_l&L
  \end{array}
  \right\rbrace
  X_L(ijlk) ,
\label{eq7}
\end{multline}
where the quantities $X_L(ijkl)$ are given by
\begin{equation}
X_L(ijkl) = (-1)^L \langle\kappa_i\|C_L\|\kappa_k\rangle
         \langle\kappa_j\|C_L\|\kappa_l\rangle R_L(ijkl) .\label{eq8}
\end{equation}
The coefficients $\langle\kappa_i\|C_L\|\kappa_j\rangle$ in the above
equation are reduced
matrix elements of normalized spherical harmonics, and the
quantities $R_L(ijkl)$ are relativistic Slater integrals
\citep{CCJ:93,PJS:94}. In the present calculation,
where both the Coulomb and Breit interactions are included in the Hamiltonian,
\begin{multline}
X_L(ijkl) \rightarrow X_L(ijkl) \\
+ M_L(ijkl) + N_L(ijkl) + O_L(ijkl) \, , \label{eq9}
\end{multline}
where $M_L(ijkl)$, $N_L(ijkl)$, and $O_L(ijkl)$ are magnetic
Slater integrals \citep{JBS:88}.

Identification of the levels obtained by solving the CI equations was aided
by comparison with the online \citet{nist:01}.

\subsection{Transition Amplitudes and Rates}
Using the CI wave functions discussed in Sec.~\ref{seca} for both the
initial and final states and carrying out the sums over magnetic substates,
one obtains the following expression for the reduced
electric-dipole matrix element
\begin{multline}
  \langle F || D|| I \rangle = -
   \sqrt{[J_I][J_F]} \sum_{\substack{m\le n\\ r\le s}}
   \eta_{rs} \, \eta_{mn} \,
   c^{\scriptscriptstyle (F)}_{rs} \,
   c^{\scriptscriptstyle (I)}_{mn} \,
   \times   \\
   \Biggl\{ \Biggr. (-1)^{j_r + j_s + J_I}
      \left\{
      \begin{array}{ccc}
      1 & J_I & J_F \\ j_s & j_r & j_m
     \end{array}
      \right \}
       \langle r || d || m \rangle  \delta_{ns} \\
 +  (-1)^{j_r + j_n}
      \left\{
      \begin{array}{ccc}
      1 & J_I & J_F\\ j_s & j_r & j_n
     \end{array}
      \right \}
     \langle r || d || n \rangle  \delta_{ms}  \\
    +   (-1)^{J_F + J_I + 1}
      \left\{
      \begin{array}{ccc}
      1 & J_I & J_F\\ j_r & j_s & j_m
     \end{array}
      \right \}
      \langle s || d || m \rangle  \delta_{nr} \\
 + (-1)^{j_r + j_n + J_F}
      \left\{
      \begin{array}{ccc}
      1 & J_I & J_F\\ j_r & j_s & j_n
     \end{array}
      \right \}
 \langle s || d || n \rangle  \delta_{mr} \Biggl. \Biggr\}
  , \label{eq10}
\end{multline}
where $[J]=2J+1$.
The one-electron reduced matrix elements $\langle m \| d \| n \rangle$
in Eq.~(\ref{eq10}) are given by
\begin{multline}
\langle\kappa_i||d||\kappa_j\rangle
 = \frac{3}{k} \langle  \kappa_i|| C_1||  \kappa_j \rangle \\
    \int_0^\infty \! \! dr \, \biggl\{ j_1(kr) [P_i(r)P_j(r) + Q_i(r)Q_j(r)]
                                   \biggr. \\
+  j_2(kr) \left[ \frac{\kappa_i - \kappa_j}{2} [P_i(r) Q_j(r)
                                  + Q_i(r) P_j(r)] \right. \\
+ [P_i(r) Q_j(r) - Q_i(r) P_j(r)] \left. \right] \biggl. \biggr\} \, ,
                                       \label{dl}
\end{multline}
in length form, and
\begin{multline}
\langle\kappa_i||d||\kappa_j\rangle
 = \frac{3}{k} \langle  \kappa_i|| C_1 ||  \kappa_j \rangle \\
\int_0^\infty \!\! dr \, \Biggl\{ \Biggr. - \frac{\kappa_i - \kappa_j}{2}
  \left [ \frac{d j_1(kr)}{d\, kr} + \frac{j_1(kr)}{kr} \right ] \times \\
[P_i(r) Q_j(r) + Q_i(r) P_j(r)]\Biggr.  \\
 \Biggl.
 +  \frac{j_1(kr)}{kr} \,
[P_i(r)Q_j(r) - Q_i(r)P_j(r)] \Biggr\} \, , \label{dv}
\end{multline}
in velocity form.
In Eqs.~(\ref{dl}) and (\ref{dv}), the quantities $P_i(r)$ and $Q_i(r)$ are
large- and small-component radial Dirac wave functions for state $i$
and $j_l(kr)$ is a spherical Bessel function of order $l$; $k=2\pi/\lambda$ being
the magnitude of the wave vector. These dipole matrix elements are fully retarded.
The dipole transition rates are given in terms of the dipole matrix elements by
\begin{equation}
A_{FI} = \frac{2.0613\times 10^{18}}{[J_F]\lambda^3}  S_{FI}, \label{trans}
\end{equation}
where $S_{FI} = |\langle F || D|| I \rangle|^2$ is the line-strength of the transition
(atomic units) and $\lambda$ is the transition wavelength (\AA).

\section{Discussion of Tables}

We solve the CI-equation (\ref{eq6}) using the method described in
\cite{CCJ:93} to obtain wave functions for $(1snl)\ ^{2S+1}\!L_J$ states
with $J \leq 3$, $S= 0$ \& 1,  and $n \leq 6$ for  helium-like ions
with $Z$ = 6, 7, 8, 10, 14, and 18.  The energies obtained from the CI
calculations are in precise agreement with earlier relativistic
calculations \citep{PJS:94} and in agreement to parts in 10$^4$ with the
nonrelativistic CI calculations of \citet{CT:92}.
The wavelengths for transitions between nearly degenerate levels
are less accurate than the level energies owing to cancellation.
The present
wavelengths agree with NIST tabulations of measured wavelengths \citep{WFD:96}
to better than
0.02\% for $\lambda < 200$~\AA, to better than 0.2\% for
$200\leq \lambda < 2000$~\AA, and to better than 2\% for $\lambda < 20000$~\AA.
Oscillator strengths from the present calculation agree precisely
with those given in \cite{PJS:95} for transitions between states with
with $n=1$ and 2; they are also in close agreement (typically 0.01\%)
with values from \cite{CT:92}.
Transition rates agree with the tabulated NIST values \citep{WFD:96}
to better than
0.5\% for $\lambda < 200$~\AA, to better than 1\% for
$200\leq \lambda < 2000$ \AA, and to better than 2\% for $\lambda < 20000$\AA.

Results of our calculations are given in
Tables~\ref{sps}--\ref{xsp}. Although we quote five
significant figures for wavelengths and four significant figures for
transition rates in these tables, the reader is cautioned that
the accuracy of the wavelengths (and consequently the transition rates)
as determined by the comparisons discussed above is substantially smaller
for larger wavelengths.  Length-form and velocity-form matrix elements
for these transitions are in close agreement; however, there are small
residual differences
between length-form and velocity-form matrix elements caused by our neglect of
contributions from negative-energy states.
As discussed in \cite{PJS:95}, these differences, when evaluated perturbatively,
contribute only to velocity-form matrix elements, and bring velocity-form matrix
elements into precise agreement with the corresponding length-form matrix elements.
In the present tabulation,  we list transition rates obtained from length-form
calculations only.

In Tables~\ref{sps}--\ref{sdf}, we present wavelengths and transition
rates for singlet-singlet transitions of the type $(1snl)\, ^1L -(1sml')\, ^1L'$
with $L$ and $L'$ ranging through $S,\, P,\, D,\, F$.  With the exception of
the $n\, ^1 P - m\, ^1 S$ transitions listed in Table~\ref{sps},
$n-n$ transitions (which have wavelengths $\sim 10^5\ \text{to}\ 10^6$ \AA\ and
transition rates $< 10^5$ s$^{-1}$) are omitted from the tables since
{\it ab-initio} calculations for such cases are unreliable.

In Tables~\ref{tps}--\ref{tpd}
we present wavelengths and transition
rates for triplet-triplet transitions of the type $(1snl)\, ^3L -(1sml')\, ^3 L'$
with $L$ and $L'$ ranging over $S,\, P,\, D$.
In the astrophysical applications mentioned in the introduction,
the fine-structure of the transitions is unresolved.
We therefore average the triplet-triplet rates over
individual fine-structure substates $J_I$, $J_F$.
The average  rates $\bar{A}$
listed in the tables are given by
\begin{equation}
\bar{A} = \frac{\sum_{J_I,J_F} [J_F] A_{FI} }{3 [L_F]} .
\end{equation}
(Note that $3 [L_F] = \sum_{J_F} [J_F]$ for triplet states.)
Multiplet-average wavelengths $\bar{\lambda}$ are also listed in the tables.
For reasons given in the previous paragraph, we include $n-n$ transition data only for the
$n\ ^3 \!P - m\ ^3\!S$ transitions (Table~\ref{tps}).

The intercombination transitions between $n\ ^3\! P_1$ states and $m\ ^1\! S_0$ states,
which are comparable in size to allowed transitions for highly-charged ions,
are listed in Tables~\ref{xps} \& \ref{xsp}.

In summary, we present accurate wavelengths and transition rates
for allowed singlet-singlet and triplet-triplet transitions
and for intercombination transitions
between triplet $^3P_1$ and singlet $^1S_0$ states with $n\leq 6$
in helium-like, carbon, nitrogen, oxygen,
neon, silicon, and argon for use in the analysis of astrophysical plasmas.

\acknowledgments
The research of WRJ and IS was supported in part by National Science Foundation
Grant No.\ PHY-99-70666. The research of UIS was supported by Grant No.\
B503968 from Lawrence Livermore National Laboratory.
The research of AD was supported by the Division of Astronomy of the National
Science Foundation.

\begin{deluxetable}{lrrrrrrrrrr}
\tablecaption{Wavelengths $\lambda$ (\AA) and radiative transition rates
$A$ (s$^{-1}$) for singlet-singlet transitions in He-like
ions. Numbers in brackets represent powers of 10.\label{sps}}
\tablecolumns{11}
\tablewidth{0pc}
\tablehead{
\multicolumn{1}{c}{Final} &
\multicolumn{2}{c}{$1\, ^1\! S_0$} &
\multicolumn{2}{c}{$2\, ^1\! S_0$} &
\multicolumn{2}{c}{$3\, ^1\! S_0$} &
\multicolumn{2}{c}{$4\, ^1\! S_0$} &
\multicolumn{2}{c}{$5\, ^1\! S_0$} \\
\multicolumn{1}{c}{Initial} &
\multicolumn{1}{c}{$\lambda$  (\AA)} &
\multicolumn{1}{c}{$A$ (s$^{-1}$)}&
\multicolumn{1}{c}{$\lambda$ (\AA)} &
\multicolumn{1}{c}{$A$ (s$^{-1}$)} &
\multicolumn{1}{c}{$\lambda$  (\AA)} &
\multicolumn{1}{c}{$A$ (s$^{-1}$)} &
\multicolumn{1}{c}{$\lambda$ (\AA)} &
\multicolumn{1}{c}{$A$ (s$^{-1}$)} &
\multicolumn{1}{c}{$\lambda$ (\AA)} &
\multicolumn{1}{c}{$A$ (s$^{-1}$)} }
\startdata
\multicolumn{11}{c}{He-like carbon}\\
$2\, ^1\! P_1$&  40.266& 8.862[11]&  3524.7& 1.672[07]&        &	  &	   &	      &        &	  \\
$3\, ^1\! P_1$&  34.972& 2.551[11]&  247.31& 1.277[10]&  12160.& 2.440[06]&	   &	      &        &	  \\
$4\, ^1\! P_1$&  33.426& 1.064[11]&  186.35& 5.764[09]&  711.69& 1.663[09]&  29083.& 5.936[05]&        &	  \\
$5\, ^1\! P_1$&  32.754& 5.420[10]&  167.22& 3.000[09]&  495.34& 9.318[08]&  1543.0& 3.897[08]&  57105.& 1.959[05]\\
$6\, ^1\! P_1$&  32.399& 3.186[10]&  158.36& 1.780[09]&  424.88& 5.638[08]&  1017.4& 2.514[08]&  2838.4& 1.288[08]\\
\multicolumn{11}{c}{He-like nitrogen} \\
$2\, ^1\! P_1$&  28.786& 1.806[12]&  2893.3& 2.098[07]&        &	  &	   &	      &        &	  \\
$3\, ^1\! P_1$&  24.900& 5.149[11]&  173.39& 2.690[10]&  9972.0& 3.072[06]&	   &	      &        &	  \\
$4\, ^1\! P_1$&  23.771& 2.141[11]&  130.30& 1.205[10]&  498.19& 3.531[09]&  23835.& 7.489[05]&        &	  \\
$5\, ^1\! P_1$&  23.281& 1.089[11]&  116.84& 6.259[09]&  345.82& 1.965[09]&  1079.5& 8.316[08]&  46755.& 2.478[05]\\
$6\, ^1\! P_1$&  23.024& 6.328[10]&  110.62& 3.667[09]&  296.47& 1.172[09]&  710.40& 5.268[08]&  1989.4& 2.721[08]\\
\multicolumn{11}{c}{He-like oxygen} \\
$2\, ^1\! P_1$&  21.600& 3.302[12]&  2446.6& 2.547[07]&        &	  &	   &	      &        &	  \\
$3\, ^1\! P_1$&  18.627& 9.344[11]&  128.24& 5.039[10]&  8427.3& 3.738[06]&	   &	      &        &	  \\
$4\, ^1\! P_1$&  17.767& 3.876[11]&  96.194& 2.246[10]&  368.10& 6.654[09]&  20139.& 9.120[05]&        &	  \\
$5\, ^1\! P_1$&  17.395& 1.970[11]&  86.204& 1.164[10]&  255.01& 3.683[09]&  797.24& 1.571[09]&  39507.& 3.018[05]\\
$6\, ^1\! P_1$&  17.199& 1.146[11]&  81.591& 6.828[09]&  218.48& 2.198[09]&  523.52& 9.925[08]&  1468.2& 5.165[08]\\
\multicolumn{11}{c}{He-like neon} \\
$2\, ^1\! P_1$&  13.446& 8.853[12]&  1852.6& 3.542[07]&        &	  &	   &	      &        &	  \\
$3\, ^1\! P_1$&  11.546& 2.478[12]&  78.251& 1.397[11]&  6375.9& 5.214[06]&	   &	      &        &	  \\
$4\, ^1\! P_1$&  10.999& 1.024[12]&  58.547& 6.182[10]&  224.32& 1.859[10]&  15231.& 1.274[06]&        &	  \\
$5\, ^1\! P_1$&  10.764& 5.196[11]&  52.427& 3.194[10]&  155.00& 1.021[10]&  485.61& 4.405[09]&  29849.& 4.228[05]\\
$6\, ^1\! P_1$&  10.639& 3.008[11]&  49.607& 1.864[10]&  132.70& 6.051[09]&  318.10& 2.752[09]&  894.55& 1.446[09]\\
\multicolumn{11}{c}{He-like silicon} \\
$2\, ^1\! P_1$&  6.6466& 3.757[13]&  1195.5& 6.276[07]&        &          &        &          &        &          \\
$3\, ^1\! P_1$&  5.6796& 1.037[13]&  37.807& 6.156[11]&  4105.5& 9.302[06]&        &          &        &          \\
$4\, ^1\! P_1$&  5.4036& 4.270[12]&  28.214& 2.703[11]&  108.26& 8.246[10]&  9801.1& 2.278[06]&        &          \\
$5\, ^1\! P_1$&  5.2846& 2.161[12]&  25.246& 1.392[11]&  74.607& 4.495[10]&  234.26& 1.961[10]&  19189.& 7.586[05]\\
$6\, ^1\! P_1$&  5.2221& 1.244[12]&  23.880& 8.076[10]&  63.822& 2.644[10]&  153.05& 1.211[10]&  431.60& 6.426[09]\\
\multicolumn{11}{c}{He-like argon} \\
$2\, ^1\! P_1$&  3.9478& 1.071[14]&  816.70& 1.133[08]&        &     &        & 	 &	  &	     \\
$3\, ^1\! P_1$&  3.3647& 2.931[13]&  22.162& 1.790[12]&  2793.9& 1.697[07]&        &          &        &          \\
$4\, ^1\! P_1$&  3.1989& 1.203[13]&  16.521& 7.826[11]&  63.436& 2.402[11]&  6661.5& 4.173[06]&        &          \\
$5\, ^1\! P_1$&  3.1275& 6.078[12]&  14.779& 4.024[11]&  43.669& 1.305[11]&  137.25& 5.717[10]&  13045.& 1.389[06]\\
$6\, ^1\! P_1$&  3.0900& 3.535[12]&  13.976& 2.358[11]&  37.335& 7.749[10]&  89.522& 3.557[10]&  252.45& 1.896[10]\\
\enddata
\end{deluxetable}

\begin{deluxetable}{lrrrrrrrrr}
\tablecaption{Wavelengths $\lambda$ (\AA) and radiative transition rates
$A$ (s$^{-1}$) for singlet-singlet transitions in He-like
ions. Numbers in brackets represent powers of 10.\label{ssp}}
\tablecolumns{8}
\tablewidth{0pc}
\tablehead{
\multicolumn{1}{c}{Final} &
\multicolumn{2}{c}{$2\, ^1\! P_1$} &
\multicolumn{2}{c}{$3\, ^1\! P_1$} &
\multicolumn{2}{c}{$4\, ^1\! P_1$} &
\multicolumn{2}{c}{$5\, ^1\! P_1$}\\
\multicolumn{1}{c}{Initial} &
\multicolumn{1}{c}{$\lambda$ (\AA)} &
\multicolumn{1}{c}{$A$ (s$^{-1}$)} &
\multicolumn{1}{c}{$\lambda$ (\AA)} &
\multicolumn{1}{c}{$A$ (s$^{-1}$)} &
\multicolumn{1}{c}{$\lambda$ (\AA)} &
\multicolumn{1}{c}{$A$ (s$^{-1}$)} &
\multicolumn{1}{c}{$\lambda$ (\AA)} &
\multicolumn{1}{c}{$A$ (s$^{-1}$)}
}
\startdata
\multicolumn{9}{c}{He-like carbon} \\
$3\, ^1\! S_0$&  271.92& 5.707[09]&	   &	      &        &	  &	   &	       \\
$4\, ^1\! S_0$&  198.09& 2.292[09]&  776.10& 1.565[09]&        &	  &	   &	       \\
$5\, ^1\! S_0$&  176.09& 1.141[09]&  521.09& 7.539[08]&  1677.3& 5.361[08]&	   &	       \\
$6\, ^1\! S_0$&  166.09& 6.613[08]&  442.23& 4.272[08]&  1065.6& 2.954[08]&  3079.7& 2.234[08] \\
\multicolumn{9}{c}{He-like nitrogen} \\
$3\, ^1\! S_0$&  187.92& 1.127[10]&	   &	      &        &	  &	   &	       \\
$4\, ^1\! S_0$&  137.23& 4.542[09]&  536.18& 3.110[09]&        &	  &	   &	       \\
$5\, ^1\! S_0$&  122.07& 2.265[09]&  361.01& 1.505[09]&  1158.7& 1.070[09]&	   &	       \\
$6\, ^1\! S_0$&  115.18& 1.301[09]&  306.71& 8.458[08]&  738.89& 5.865[08]&  2132.4& 4.418[08] \\
\multicolumn{9}{c}{He-like oxygen} \\
$3\, ^1\! S_0$&  137.55& 2.013[10]&	   &	      &        &	  &	   &	       \\
$4\, ^1\! S_0$&  100.63& 8.131[09]&  392.41& 5.583[09]&        &	  &	   &	       \\
$5\, ^1\! S_0$&  89.555& 4.057[09]&  264.73& 2.708[09]&  847.91& 1.925[09]&	   &	       \\
$6\, ^1\! S_0$&  84.510& 2.336[09]&  225.03& 1.527[09]&  541.73& 1.060[09]&  1559.5& 7.983[08] \\
\multicolumn{9}{c}{He-like neon} \\
$3\, ^1\! S_0$&  82.762& 5.227[10]&	   &	      &        &	  &	   &	       \\
$4\, ^1\! S_0$&  60.698& 2.117[10]&  236.11& 1.460[10]&        &	  &	   &	       \\
$5\, ^1\! S_0$&  54.052& 1.057[10]&  159.72& 7.105[09]&  510.17& 5.052[09]&	   &	       \\
$6\, ^1\! S_0$&  51.022& 6.066[09]&  135.88& 3.995[09]&  326.94& 2.783[09]&  938.91& 2.094[09] \\
\multicolumn{9}{c}{He-like silicon} \\
$3\, ^1\! S_0$&  39.416& 2.149[11]&	   &	      &        &	  &	   &	       \\
$4\, ^1\! S_0$&  28.981& 8.726[10]&  112.47& 6.054[10]&        &	  &	   &	       \\
$5\, ^1\! S_0$&  25.825& 4.356[10]&  76.290& 2.954[10]&  243.03& 2.100[10]&	   &	       \\
$6\, ^1\! S_0$&  24.385& 2.490[10]&  64.957& 1.655[10]&  156.22& 1.157[10]&  447.49& 8.696[09] \\
\multicolumn{9}{c}{He-like argon} \\
$3\, ^1\! S_0$&  22.967& 6.071[11]&	  &	     &        & 	 &	  &	      \\
$4\, ^1\! S_0$&  16.905& 2.467[11]&  65.549& 1.715[11]&        & 	 &	  &	      \\
$5\, ^1\! S_0$&  15.069& 1.232[11]&  44.514& 8.380[10]&  141.66& 5.956[10]&	  &	      \\
$6\, ^1\! S_0$&  14.229& 7.123[10]&  37.903& 4.753[10]&  91.101& 3.323[10]&  260.35& 2.495[10] \\
\enddata
\end{deluxetable}

\begin{deluxetable}{lrrrrrrrrr}
\tablecaption{Wavelengths $\lambda$ (\AA) and radiative transition rates
$A$ (s$^{-1}$) for singlet-singlet transitions in He-like
ions. Numbers in brackets represent powers of 10.\label{sdp}}
\tablecolumns{8}
\tablewidth{0pc}
\tablehead{
\multicolumn{1}{c}{Final} &
\multicolumn{2}{c}{$2\, ^1\! P_1$} &
\multicolumn{2}{c}{$3\, ^1\! P_1$} &
\multicolumn{2}{c}{$4\, ^1\! P_1$} &
\multicolumn{2}{c}{$5\, ^1\! P_1$}\\
\multicolumn{1}{c}{Initial} &
\multicolumn{1}{c}{$\lambda$ (\AA)} &
\multicolumn{1}{c}{$A$ (s$^{-1}$)} &
\multicolumn{1}{c}{$\lambda$ (\AA)} &
\multicolumn{1}{c}{$A$ (s$^{-1}$)} &
\multicolumn{1}{c}{$\lambda$ (\AA)} &
\multicolumn{1}{c}{$A$ (s$^{-1}$)} &
\multicolumn{1}{c}{$\lambda$ (\AA)} &
\multicolumn{1}{c}{$A$ (s$^{-1}$)}
}
\startdata
\multicolumn{9}{c}{He-like carbon} \\
$3\, ^1\! D_2$&  267.28& 3.930[10]&	   &	      &        &	  &	   &	       \\
$4\, ^1\! D_2$&  197.03& 1.229[10]&  760.16& 4.410[09]&        &	  &	   &	       \\
$5\, ^1\! D_2$&  175.67& 5.567[09]&  517.36& 2.096[09]&  1639.3& 9.415[08]&	   &	       \\
$6\, ^1\! D_2$&  165.88& 3.062[09]&  440.75& 1.169[09]&  1057.1& 5.468[08]&  3009.5& 2.905[08] \\
\multicolumn{9}{c}{He-like nitrogen} \\
$3\, ^1\! D_2$&  185.22& 8.137[10]&	   &	      &        &	  &	   &	       \\
$4\, ^1\! D_2$&  136.62& 2.555[10]&  526.87& 9.117[09]&        &	  &	   &	       \\
$5\, ^1\! D_2$&  121.82& 1.159[10]&  358.81& 4.344[09]&  1136.4& 1.944[09]&	   &	       \\
$6\, ^1\! D_2$&  115.05& 6.319[09]&  305.80& 2.401[09]&  733.62& 1.121[09]&  2089.1& 5.934[08] \\
\multicolumn{9}{c}{He-like oxygen} \\
$3\, ^1\! D_2$&  135.83& 1.501[11]&	   &	      &        &	  &	   &	       \\
$4\, ^1\! D_2$&  100.23& 4.734[10]&  386.47& 1.682[10]&        &	  &	   &	       \\
$5\, ^1\! D_2$&  89.394& 2.151[10]&  263.33& 8.033[09]&  833.67& 3.585[09]&	   &	       \\
$6\, ^1\! D_2$&  84.429& 1.176[10]&  224.45& 4.454[09]&  538.42& 2.075[09]&  1532.3& 1.096[09] \\
\multicolumn{9}{c}{He-like neon} \\
$3\, ^1\! D_2$&  81.925& 4.025[11]&	   &	      &        &	  &	   &	       \\
$4\, ^1\! D_2$&  60.504& 1.284[11]&  233.20& 4.529[10]&        &	  &	   &	       \\
$5\, ^1\! D_2$&  53.972& 5.853[10]&  159.02& 2.173[10]&  503.17& 9.658[09]&	   &	       \\
$6\, ^1\! D_2$&  50.981& 3.199[10]&  135.59& 1.205[10]&  325.26& 5.593[09]&  925.25& 2.945[09] \\
\multicolumn{9}{c}{He-like silicon} \\
$3\, ^1\! D_2$&  39.098& 1.628[12]&	   &	      &        &	  &	   &	       \\
$4\, ^1\! D_2$&  28.907& 5.316[11]&  111.37& 1.860[11]&        &	  &	   &	       \\
$5\, ^1\! D_2$&  25.795& 2.445[11]&  76.026& 9.014[10]&  240.37& 3.985[10]&	   &	       \\
$6\, ^1\! D_2$&  24.369& 1.339[11]&  64.845& 5.010[10]&  155.57& 2.318[10]&  442.21& 1.215[10] \\
\multicolumn{9}{c}{He-like argon} \\
$3\, ^1\! D_2$&  22.791& 4.502[12]&	   &	      &        &	  &	   &	       \\
$4\, ^1\! D_2$&  16.864& 1.475[12]&  64.938& 5.142[11]&        &	  &	   &	       \\
$5\, ^1\! D_2$&  15.052& 6.801[11]&  44.368& 2.501[11]&  140.19& 1.102[11]&	   &	       \\
$6\, ^1\! D_2$&  14.220& 3.761[11]&  37.846& 1.403[11]&  90.770& 6.481[10]&  257.67& 3.390[10] \\
\enddata
\end{deluxetable}

\begin{deluxetable}{crrrrrr}
\tablecaption{Wavelengths $\lambda$ (\AA) and radiative transition rates
$A$ (s$^{-1}$) for singlet-singlet transitions in He-like
ions. Numbers in brackets represent powers of 10.\label{spd}}
\tablehead{
\multicolumn{1}{c}{Final} &
\multicolumn{2}{c}{$3\, ^1\! D_2$} &
\multicolumn{2}{c}{$4\, ^1\! D_2$} &
\multicolumn{2}{c}{$5\, ^1\! D_2$} \\
\multicolumn{1}{c}{Initial} &
\multicolumn{1}{c}{$\lambda$ (\AA)}&
\multicolumn{1}{c}{$A$ (s$^{-1}$)} &
\multicolumn{1}{c}{$\lambda$ (\AA)} &
\multicolumn{1}{c}{$A$ (s$^{-1}$)} &
\multicolumn{1}{c}{$\lambda$ (\AA)} &
\multicolumn{1}{c}{$A$ (s$^{-1}$)}  }
\startdata
\multicolumn{7}{c}{He-like carbon} \\
$4\, ^1\! P_1$&  745.56& 1.953[08]&	  &	     &        & 	 \\
$5\, ^1\! P_1$&  511.51& 8.443[07]& 1610.1& 1.077[08]&        & 	 \\
$6\, ^1\! P_1$&  436.72& 4.505[07]& 1046.2& 5.503[07]&  2954.2& 5.631[07]\\
\multicolumn{7}{c}{He-like nitrogen} \\
$4\, ^1\! P_1$&  518.22& 4.100[08]&	  &	     &        & 	 \\
$5\, ^1\! P_1$&  355.35& 1.773[08]& 1119.3& 2.263[08]&        & 	 \\
$6\, ^1\! P_1$&  303.45& 9.376[07]& 727.44& 1.145[08]&  2058.9& 1.170[08]\\
\multicolumn{7}{c}{He-like oxygen} \\
$4\, ^1\! P_1$&  381.00& 7.648[08]&	   &	      &       & 	 \\
$5\, ^1\! P_1$&  261.14& 3.306[08]&  822.94& 4.223[08]&       & 	 \\
$6\, ^1\! P_1$&  222.96& 1.752[08]&  534.49& 2.141[08]& 1513.0& 2.188[08]\\
\multicolumn{7}{c}{He-like neon} \\
$4\, ^1\! P_1$&  230.72& 2.085[09]&	   &	      &       & 	 \\
$5\, ^1\! P_1$&  158.03& 9.007[08]&  498.38& 1.158[09]&       & 	 \\
$6\, ^1\! P_1$&  134.91& 4.761[08]&  323.53& 5.851[08]& 916.91& 5.995[08]\\
\multicolumn{7}{c}{He-like silicon} \\
$4\, ^1\! P_1$&  110.73& 8.650[09]&	  &	     &        & 	 \\
$5\, ^1\! P_1$&  75.772& 3.734[09]& 239.19& 4.874[09]&        & 	 \\
$6\, ^1\! P_1$&  64.673& 1.967[09]& 155.15& 2.454[09]&  440.25& 2.530[09]\\
\multicolumn{7}{c}{He-like argon} \\
$4\, ^1\! P_1$&  64.821& 2.448[10]&	  &	     &        & 	 \\
$5\, ^1\! P_1$&  44.321& 1.057[10]& 140.01& 1.375[10]&        & 	 \\
$6\, ^1\! P_1$&  37.811& 5.621[09]& 90.688& 6.985[09]&  257.25& 7.201[09]\\
\enddata
\end{deluxetable}

\begin{deluxetable}{crrrrrr}
\tablecaption{Wavelengths $\lambda$ (\AA) and radiative transition rates
$A$ (s$^{-1}$) for singlet-singlet transitions in He-like
ions. Numbers in brackets represent powers of 10.\label{sfd}}
\tablehead{
\multicolumn{1}{c}{Final} &
\multicolumn{2}{c}{$3\, ^1\! D_2$} &
\multicolumn{2}{c}{$4\, ^1\! D_2$} &
\multicolumn{2}{c}{$5\, ^1\! D_2$} \\
\multicolumn{1}{c}{Initial} &
\multicolumn{1}{c}{$\lambda$ (\AA)}&
\multicolumn{1}{c}{$A$ (s$^{-1}$)} &
\multicolumn{1}{c}{$\lambda$ (\AA)} &
\multicolumn{1}{c}{$A$ (s$^{-1}$)} &
\multicolumn{1}{c}{$\lambda$ (\AA)} &
\multicolumn{1}{c}{$A$ (s$^{-1}$)}  }
\startdata
\multicolumn{7}{c}{He-like carbon} \\
$4\, ^1\! F_3$&  749.60& 6.005[09]&	  &	     &        & 	 \\
$5\, ^1\! F_3$&  512.46& 2.101[09]& 1619.6& 1.176[09]&        & 	 \\
$6\, ^1\! F_3$&  437.25& 1.026[09]& 1049.2& 6.061[08]&  2978.7& 3.391[08]\\
\multicolumn{7}{c}{He-like nitrogen} \\
$4\, ^1\! F_3$&  520.61& 1.255[10]&	  &	     &        & 	 \\
$5\, ^1\! F_3$&  355.90& 4.329[09]& 1124.8& 2.407[09]&        & 	 \\
$6\, ^1\! F_3$&  303.70& 2.089[09]& 728.86& 1.226[09]&  2070.3& 6.846[08]\\
\multicolumn{7}{c}{He-like oxygen} \\
$4\, ^1\! F_3$&  382.50& 2.377[10]&	   &	      &       & 	 \\
$5\, ^1\! F_3$&  261.49& 8.103[09]&  826.41& 4.477[09]&       & 	 \\
$6\, ^1\! F_3$&  223.12& 3.906[09]&  535.46& 2.278[09]& 1520.8& 1.269[09]\\
\multicolumn{7}{c}{He-like neon} \\
$4\, ^1\! F_3$&  231.38& 6.952[10]&	   &	      &       & 	 \\
$5\, ^1\! F_3$&  158.18& 2.333[10]&  499.88& 1.276[10]&       & 	 \\
$6\, ^1\! F_3$&  134.98& 1.115[10]&  323.92& 6.436[09]& 920.05& 3.569[09]\\
\multicolumn{7}{c}{He-like silicon} \\
$4\, ^1\! F_3$&  110.88& 3.441[11]&	  &	     &        & 	 \\
$5\, ^1\! F_3$&  75.803& 1.140[11]& 239.51& 6.276[10]&        & 	 \\
$6\, ^1\! F_3$&  64.687& 5.402[10]& 155.22& 3.138[10]& 440.88 & 1.740[10]\\
\multicolumn{7}{c}{He-like argon} \\
$4\, ^1\! F_3$&  64.821& 1.066[12]&	  &	     &        & 	 \\
$5\, ^1\! F_3$&  44.319& 3.517[11]& 140.00& 1.972[11]&        & 	 \\
$6\, ^1\! F_3$&  37.817& 1.671[11]& 90.721& 9.884[10]& 257.52 & 5.514[10]\\
\enddata
\end{deluxetable}

\begin{deluxetable}{lrrrr}
\tablecaption{Wavelengths $\lambda$ (\AA) and radiative transition rates
$A$ (s$^{-1}$) for singlet-singlet transitions in He-like
ions. Numbers in brackets represent powers of 10.\label{sdf}}
\tablehead{
\multicolumn{1}{c}{Final} &
\multicolumn{2}{c}{$4\, ^1\! F_2$} &
\multicolumn{2}{c}{$5\, ^1\! F_3$} \\
\multicolumn{1}{c}{Initial} &
\multicolumn{1}{c}{$\lambda$ (\AA)} &
\multicolumn{1}{c}{$A$ (s$^{-1}$)} &
\multicolumn{1}{c}{$\lambda$ (\AA)} &
\multicolumn{1}{c}{$A$ (s$^{-1}$)} }
\startdata
\multicolumn{5}{c}{He-like carbon} \\
$ 5\, ^1\! D_2$&     1620.0&   2.149[07]&	  &	       \\
$ 6\, ^1\! D_2$&     1049.0&   9.238[06]&    2976.6& 1.794[07]  \\
\multicolumn{5}{c}{He-like nitrogen} \\
$ 5\, ^1\! D_2$&     1125.0&   4.449[07]&	  &	       \\
$ 6\, ^1\! D_2$&     728.87&   1.893[07]&    2070.1&   3.623[07]\\
\multicolumn{5}{c}{He-like oxygen} \\
$ 5\, ^1\! D_2$&     826.54&   8.348[07]&	  &	       \\
$ 6\, ^1\! D_2$&     535.44&   3.553[07]&    1520.4&   6.723[07]\\
\multicolumn{5}{c}{He-like neon} \\
$ 5\, ^1\! D_2$&     500.03&   2.411[08]&	  &	       \\
$ 6\, ^1\! D_2$&     323.95&   1.022[08]&    920.08&   1.902[08]\\
\multicolumn{5}{c}{He-like silicon} \\
$ 5\, ^1\! D_2$&     239.70&   1.207[09]&	  &	       \\
$ 6\, ^1\! D_2$&     155.29&   5.101[08]&    441.19&   9.351[08]\\
\multicolumn{5}{c}{He-like argon} \\
$ 5\, ^1\! D_2$&     140.19&   3.850[09]&	  &	       \\
$ 6\, ^1\! D_2$&     90.771&   1.644[09]&    257.71&   3.000[09]\\
\enddata
\end{deluxetable}

\begin{deluxetable}{lrrrrrrrr}
\tablecaption{Average wavelengths $\bar{\lambda}$ (\AA) and radiative
transition rates $\bar{A}$ (s$^{-1}$) for triplet-triplet transitions
in He-like ions. Numbers in brackets represent powers of 10.\label{tps}}

\tablehead{
\multicolumn{1}{c}{Final} &
\multicolumn{2}{c}{$2\, ^3\! S$} &
\multicolumn{2}{c}{$3\, ^3\!S$} &
\multicolumn{2}{c}{$4\, ^3\!S$} &
\multicolumn{2}{c}{$5\, ^3\!S$} \\
\multicolumn{1}{c}{Initial} &
\multicolumn{1}{c}{$\bar{\lambda}$ (\AA)} &
\multicolumn{1}{c}{$\bar{A}$ (s$^{-1}$)} &
\multicolumn{1}{c}{$\bar{\lambda}$ (\AA)} &
\multicolumn{1}{c}{$\bar{A}$ (s$^{-1}$)} &
\multicolumn{1}{c}{$\bar{\lambda}$ (\AA)} &
\multicolumn{1}{c}{$\bar{A}$ (s$^{-1}$)} &
\multicolumn{1}{c}{$\bar{\lambda}$ (\AA)} &
\multicolumn{1}{c}{$\bar{A}$ (s$^{-1}$)} }
\startdata
\multicolumn{9}{c}{He-like carbon} \\
$2\, ^3\! P$&  2272.3 & 5.695[07] &       &          &        &          &         &         \\
$3\, ^3\! P$&  227.17 & 1.360[10] &8425.8 & 6.928[06]&        &          &         &         \\
$4\, ^3\! P$&  173.26 & 6.187[09] &671.98 & 1.673[09]& 20665. & 1.586[06]&         &         \\
$5\, ^3\! P$&  156.22 & 3.223[09] &472.13 & 9.483[08]& 1474.3 & 3.821[08]& 41075.  &5.087[05]\\
$6\, ^3\! P$&  148.30 & 1.906[09] &406.54 & 5.738[08]& 980.37 & 2.493[08]& 2731.8  &1.240[08]\\
\multicolumn{9}{c}{He-like nitrogen} \\
$2\, ^3\! P$&  1899.1 & 6.865[07] &	  &	     &        & 	 &	   &	     \\
$3\, ^3\! P$&  161.21 & 2.851[10] &7004.7 & 8.445[06]&        & 	 &	   &	     \\
$4\, ^3\! P$&  122.42 & 1.285[10] &474.41 & 3.565[09]& 17138. & 1.943[06]&	   &	     \\
$5\, ^3\! P$&  110.22 & 6.676[09] &331.95 & 2.001[09]& 1038.5 & 8.214[08]& 33998.  &6.261[05]\\
$6\, ^3\! P$&  104.57 & 3.907[09] &285.51 & 1.196[09]& 688.27 & 5.255[08]& 1925.3  &2.652[08]\\
\multicolumn{9}{c}{He-like oxygen} \\
$2\, ^3\! P$&  1627.7 & 8.085[07] &	  &	     &        & 	 &	   &	     \\
$3\, ^3\! P$&  120.32 & 5.315[10] &5979.5 & 1.003[07]&        & 	 &	   &	     \\
$4\, ^3\! P$&  91.085 & 2.380[10] &352.73 & 6.726[09]& 14605. & 2.318[06]&	   &	     \\
$5\, ^3\! P$&  81.916 & 1.234[10] &246.07 & 3.749[09]& 770.85 & 1.559[09]& 28950.  &7.480[05]\\
$6\, ^3\! P$&  77.677 & 7.233[09] &211.42 & 2.239[09]& 509.30 & 9.926[08]& 1427.1  &5.063[08]\\
\multicolumn{9}{c}{He-like neon} \\
$2\, ^3\! P$&  1256.6 & 1.077[08] &	  &	     &        & 	 &	   &	     \\
$3\, ^3\! P$&  74.350 & 1.462[11] &4588.0 & 1.354[07]&        & 	 &	   &	     \\
$4\, ^3\! P$&  56.042 & 6.493[10] &216.83 & 1.880[10]& 11179. & 3.147[06]&	   &	     \\
$5\, ^3\! P$&  50.328 & 3.356[10] &150.65 & 1.038[10]& 472.77 & 4.389[09]& 22119.  &1.020[06]\\
$6\, ^3\! P$&  47.692 & 1.957[10] &129.27 & 6.156[09]& 311.20 & 2.759[09]& 874.56  &1.428[09]\\
\multicolumn{9}{c}{He-like silicon} \\
$2\, ^3\! P$&  833.99 & 1.790[08] &	  &	     &        & 	 &	   &	     \\
$3\, ^3\! P$&  36.439 & 6.386[11] &3019.9 & 2.298[07]&        & 	 &	   &	     \\
$4\, ^3\! P$&  27.340 & 2.812[11] &105.66 & 8.348[10]& 7335.0 & 5.380[06]&	   &	     \\
$5\, ^3\! P$&  24.515 & 1.449[11] &73.103 & 4.568[10]& 229.82 & 1.964[10]& 14486.  &1.752[06]\\
$6\, ^3\! P$&  23.215 & 8.408[10] &62.637 & 2.690[10]& 150.67 & 1.219[10]& 424.68  &6.401[09]\\
\multicolumn{9}{c}{He-like argon} \\
$2\, ^3\! P$&  590.47 & 2.990[08] &	  &	     &        & 	 &	   &	     \\
$3\, ^3\! P$&  21.530 & 1.862[12] &2123.9 & 3.911[07]&        & 	 &	   &	     \\
$4\, ^3\! P$&  16.118 & 8.165[11] &62.242 & 2.453[11]& 5146.8 & 9.216[06]&	   &	     \\
$5\, ^3\! P$&  14.442 & 4.201[11] &43.087 & 1.337[11]& 135.22 & 5.794[10]& 10158.  &3.006[06]\\
$6\, ^3\! P$&  13.670 & 2.462[11] &36.792 & 7.943[10]& 88.436 & 3.617[10]& 249.32  &1.912[10]\\
\enddata
\end{deluxetable}

\begin{deluxetable}{lrrrrrrrr}
\tablecaption{Average wavelengths $\bar{\lambda}$ (\AA) and radiative
transition rates $\bar{A}$ (s$^{-1}$) for triplet-triplet transitions
in He-like ions. Numbers in brackets represent powers of 10.\label{tsp}}
\tablehead{
\multicolumn{1}{c}{Final} &
\multicolumn{2}{c}{$2\, ^3\! P$} &
\multicolumn{2}{c}{$3\, ^3\! P$} &
\multicolumn{2}{c}{$4\, ^3\! P$} &
\multicolumn{2}{c}{$5\, ^3\! P$} \\
\multicolumn{1}{c}{Initial} &
\multicolumn{1}{c}{$\bar{\lambda}$ (\AA)} &
\multicolumn{1}{c}{$\bar{A}$ (s$^{-1}$)} &
\multicolumn{1}{c}{$\bar{\lambda}$ (\AA)} &
\multicolumn{1}{c}{$\bar{A}$ (s$^{-1}$)} &
\multicolumn{1}{c}{$\bar{\lambda}$ (\AA)} &
\multicolumn{1}{c}{$\bar{A}$ (s$^{-1}$)} &
\multicolumn{1}{c}{$\bar{\lambda}$ (\AA)} &
\multicolumn{1}{c}{$\bar{A}$ (s$^{-1}$)} }
\startdata
\multicolumn{9}{c}{He-like carbon} \\
$3\, ^3\! S$& 260.20 & 6.709[09] &	 &	     &       &  	  &	   &	      \\
$4\, ^3\! S$& 189.29 & 2.607[09] & 756.96 & 1.803[09] &       &  	  &	   &	      \\
$5\, ^3\! S$& 168.44 & 1.279[09] & 506.32 & 8.527[08] & 1651.3 & 6.077[08] &	   &	      \\
$6\, ^3\! S$& 159.01 & 7.324[08] & 429.72 & 4.775[08] & 1044.2 & 3.291[08] & 3051.3 & 2.495[08]\\
\multicolumn{9}{c}{He-like nitrogen} \\
$3\, ^3\! S$& 180.71 & 1.292[10] &	  &	      &        &	   &	    &	       \\
$4\, ^3\! S$& 131.87 & 5.057[09] & 524.44 & 3.510[09] &        &	   &	    &	       \\
$5\, ^3\! S$& 117.41 & 2.488[09] & 352.07 & 1.671[09] & 1142.6 & 1.191[09] &	    &	       \\
$6\, ^3\! S$& 110.87 & 1.416[09] & 299.15 & 9.307[08] & 725.85 & 6.441[08] & 2113.7 & 4.867[08]\\
\multicolumn{9}{c}{He-like oxygen} \\
$3\, ^3\! S$& 132.81 & 2.265[10] &	 &	      &        &	   &	    &	       \\
$4\, ^3\! S$& 97.125 & 8.917[09] & 384.72 & 6.207[09] &        &	   &	    &	       \\
$5\, ^3\! S$& 86.515 & 4.395[09] & 258.93 & 2.967[09] & 837.34 & 2.114[09] &	    &	       \\
$6\, ^3\! S$& 81.702 & 2.508[09] & 220.12 & 1.659[09] & 533.29 & 1.151[09] & 1547.1 & 8.686[08]\\
\multicolumn{9}{c}{He-like neon} \\
$3\, ^3\! S$& 80.410 & 5.742[10] &	  &	      &        &	   &	    &	       \\
$4\, ^3\! S$& 58.967 & 2.275[10] & 232.31 & 1.590[10] &        &	   &	    &	       \\
$5\, ^3\! S$& 52.552 & 1.124[10] & 156.87 & 7.647[09] & 504.93 & 5.450[09] &	    &	       \\
$6\, ^3\! S$& 49.637 & 6.407[09] & 133.47 & 4.272[09] & 322.80 & 2.974[09] & 932.62 & 2.242[09]\\
\multicolumn{9}{c}{He-like silicon} \\
$3\, ^3\! S$& 38.591 & 2.303[11] &	  &	      &        &	   &	    &	       \\
$4\, ^3\! S$& 28.376 & 9.200[10] & 111.15 & 6.457[10] &        &	   &	    &	       \\
$5\, ^3\! S$& 25.302 & 4.558[10] & 75.306 & 3.124[10] & 241.20 & 2.226[10] &	    &	       \\
$6\, ^3\! S$& 23.902 & 2.592[10] & 64.127 & 1.742[10] & 154.78 & 1.217[10] & 445.25 & 9.168[09]\\
\multicolumn{9}{c}{He-like argon} \\
$3\, ^3\! S$& 22.583 & 6.463[11] &	 &	      &        &	   &	    &	       \\
$4\, ^3\! S$& 16.624 & 2.591[11] & 64.930 & 1.822[11] &        &	   &	    &	       \\
$5\, ^3\! S$& 14.825 & 1.286[11] & 44.057 & 8.840[10] & 140.79 & 6.297[10] &	    &	       \\
$6\, ^3\! S$& 14.005 & 7.402[10] & 37.519 & 4.991[10] & 90.438 & 3.491[10] & 259.31 & 2.627[10]\\
\enddata
\end{deluxetable}

\begin{deluxetable}{lrrrrrrrr}
\tablecaption{Average wavelengths $\bar{\lambda}$ (\AA) and radiative
transition rates $\bar{A}$ (s$^{-1}$) for triplet-triplet transitions
in He-like ions. Numbers in brackets represent powers of 10.\label{tdp}}
\tablehead{
\multicolumn{1}{c}{Final} &
\multicolumn{2}{c}{$2\, ^3\! P$} &
\multicolumn{2}{c}{$3\, ^3\! P$} &
\multicolumn{2}{c}{$4\, ^3\! P$} &
\multicolumn{2}{c}{$5\, ^3\! P$} \\
\multicolumn{1}{c}{Initial} &
\multicolumn{1}{c}{$\bar{\lambda}$ (\AA)} &
\multicolumn{1}{c}{$\bar{A}$ (s$^{-1}$)} &
\multicolumn{1}{c}{$\bar{\lambda}$ (\AA)} &
\multicolumn{1}{c}{$\bar{A}$ (s$^{-1}$)} &
\multicolumn{1}{c}{$\bar{\lambda}$ (\AA)} &
\multicolumn{1}{c}{$\bar{A}$ (s$^{-1}$)} &
\multicolumn{1}{c}{$\bar{\lambda}$ (\AA)} &
\multicolumn{1}{c}{$\bar{A}$ (s$^{-1}$)} }
\startdata
\multicolumn{9}{c}{He-like carbon} \\
$3\, ^3\! D$& 248.72 & 4.240[10] &	  &	      &        &	   &	    &	       \\
$4\, ^3\! D$& 186.72 & 1.412[10] & 717.49 & 4.312[09] &        &	   &	    &	       \\
$5\, ^3\! D$& 167.40 & 6.554[09] & 497.10 & 2.162[09] & 1557.2 & 8.794[08] &	    &	       \\
$6\, ^3\! D$& 158.48 & 3.650[09] & 425.91 & 1.231[09] & 1022.0 & 5.370[08] & 2869.1 & 2.640[08]\\
\multicolumn{9}{c}{He-like nitrogen} \\
$3\, ^3\! D$& 173.92 & 8.729[10] &	  &	      &        &	   &	    &	       \\
$4\, ^3\! D$& 130.35 & 2.892[10] & 501.17 & 8.966[09] &        &	   &	    &	       \\
$5\, ^3\! D$& 116.80 & 1.340[10] & 346.61 & 4.472[09] & 1087.1 & 1.838[09] &	    &	       \\
$6\, ^3\! D$& 110.56 & 7.387[09] & 296.87 & 2.518[09] & 712.56 & 1.106[09] & 2004.8 & 5.476[08]\\
\multicolumn{9}{c}{He-like oxygen} \\
$3\, ^3\! D$& 128.46 & 1.606[11] &	  &	      &        &	   &	    &	       \\
$4\, ^3\! D$& 96.148 & 5.304[10] & 369.83 & 1.663[10] &        &	   &	    &	       \\
$5\, ^3\! D$& 86.121 & 2.454[10] & 255.43 & 8.265[09] & 801.82 & 3.422[09] &	    &	       \\
$6\, ^3\! D$& 81.500 & 1.355[10] & 218.67 & 4.657[09] & 524.82 & 2.056[09] & 1478.0 & 1.025[09]\\
\multicolumn{9}{c}{He-like neon} \\
$3\, ^3\! D$& 78.294 & 4.328[11] &	  &	      &        &	   &	    &	       \\
$4\, ^3\! D$& 58.491 & 1.424[11] & 225.09 & 4.538[10] &        &	   &	    &	       \\
$5\, ^3\! D$& 52.360 & 6.580[10] & 155.17 & 2.244[10] & 487.70 & 9.392[09] &	    &	       \\
$6\, ^3\! D$& 49.538 & 3.622[10] & 132.76 & 1.259[10] & 318.65 & 5.599[09] & 898.87 & 2.815[09]\\
\multicolumn{9}{c}{He-like silicon} \\
$3\, ^3\! D$& 37.835 & 1.823[12] &	  &	      &        &	   &	    &	       \\
$4\, ^3\! D$& 28.206 & 5.994[11] & 108.57 & 1.947[11] &        &	   &	    &	       \\
$5\, ^3\! D$& 25.233 & 2.768[11] & 74.698 & 9.584[10] & 235.08 & 4.058[10] &	    &	       \\
$6\, ^3\! D$& 23.866 & 1.521[11] & 63.871 & 5.361[10] & 153.30 & 2.404[10] & 433.20 & 1.219[10]\\
\multicolumn{9}{c}{He-like argon} \\
$3\, ^3\! D$& 22.209 & 5.216[12] &	  &	      &        &	   &	    &	       \\
$4\, ^3\! D$& 16.539 & 1.708[12] & 63.659 & 5.604[11] &        &	   &	    &	       \\
$5\, ^3\! D$& 14.791 & 7.877[11] & 43.756 & 2.752[11] & 137.77 & 1.172[11] &	    &	       \\
$6\, ^3\! D$& 13.987 & 4.358[11] & 37.397 & 1.549[11] & 89.728 & 6.979[10] & 253.57 & 3.556[10]
\enddata
\end{deluxetable}

\begin{deluxetable}{lrrrrrr}
\tablecaption{Average wavelengths $\bar{\lambda}$ (\AA) and radiative
transition rates $\bar{A}$ (s$^{-1}$) for triplet-triplet transitions
in He-like ions. Numbers in brackets represent powers of 10.\label{tpd}}
\tablehead{
\multicolumn{1}{c}{Final} &
\multicolumn{2}{c}{$3\, ^3\! D$} &
\multicolumn{2}{c}{$4\, ^3\! D$} &
\multicolumn{2}{c}{$5\, ^3\! D$} \\
\multicolumn{1}{c}{Initial} &
\multicolumn{1}{c}{$\bar{\lambda}$ (\AA)}&
\multicolumn{1}{c}{$\bar{A}$ (s$^{-1}$)} &
\multicolumn{1}{c}{$\bar{\lambda}$ (\AA)} &
\multicolumn{1}{c}{$\bar{A}$ (s$^{-1}$)} &
\multicolumn{1}{c}{$\bar{\lambda}$ (\AA)} &
\multicolumn{1}{c}{$\bar{A}$ (s$^{-1}$)}}
\startdata
\multicolumn{7}{c}{He-like carbon} \\
$4\, ^3\! P$& 762.96 & 2.928[08]&	 &	    &	     &  	\\
$5\, ^3\! P$& 515.31 & 1.248[08]& 1651.1 & 1.530[08]&	     &  	\\
$6\, ^3\! P$& 438.15 & 6.627[07]& 1055.6 & 7.679[07]& 3035.4 & 7.732[07]\\
\multicolumn{7}{c}{He-like nitrogen} \\
$4\, ^3\! P$& 528.55 & 5.841[08]&	 &	    &	     &  	\\
$5\, ^3\! P$& 357.58 & 2.495[08]& 1143.6 & 3.069[08]&	     &  	\\
$6\, ^3\! P$& 304.27 & 1.313[08]& 732.93 & 1.531[08]& 2106.8 & 1.543[08]\\
\multicolumn{7}{c}{He-like oxygen} \\
$4\, ^3\! P$& 387.61 & 1.049[09]&	 &	    &	     &  	\\
$5\, ^3\! P$& 262.56 & 4.486[08]& 838.50 & 5.537[08]&	     &  	\\
$6\, ^3\! P$& 223.47 & 2.367[08]& 537.98 & 2.770[08]& 1543.6 & 2.795[08]\\
\multicolumn{7}{c}{He-like neon} \\
$4\, ^3\! P$& 233.88 & 2.734[09]&	 &	    &	     &  	\\
$5\, ^3\! P$& 158.69 & 1.170[09]& 505.81 & 1.454[09]&	     &  	\\
$6\, ^3\! P$& 135.14 & 6.155[08]& 325.18 & 7.260[08]& 931.55 & 7.350[08]\\
\multicolumn{7}{c}{He-like silicon} \\
$4\, ^3\! P$& 111.78 & 1.112[10]&	 &	    &	     &  	\\
$5\, ^3\! P$& 75.981 & 4.765[09]& 241.66 & 5.979[09]&	     &  	\\
$6\, ^3\! P$& 64.738 & 2.499[09]& 155.67 & 2.981[09]& 445.11 & 3.031[09]\\
\multicolumn{7}{c}{He-like argon} \\
$4\, ^3\! P$& 65.268 & 3.139[10]&	 &	    &	     &  	\\
$5\, ^3\! P$& 44.400 & 1.346[10]& 141.08 & 1.691[10]&	     &  	\\
$6\, ^3\! P$& 37.829 & 7.135[09]& 90.907 & 8.522[09]& 259.41 & 8.671[09]\\
\enddata
\end{deluxetable}

\begin{deluxetable}{lrrrrrrrrrr}
\tablecaption{Wavelengths $\lambda$ (\AA) and radiative
transition rates $A$ (s$^{-1}$) for triplet-singlet transitions
in He-like ions. Numbers in brackets represent powers of 10.\label{xps}}
\tablehead{
\multicolumn{1}{c}{Final} &
\multicolumn{2}{c}{$1\, ^1\! S_0$} &
\multicolumn{2}{c}{$2\, ^1\! S_0$} &
\multicolumn{2}{c}{$3\, ^1\! S_0$} &
\multicolumn{2}{c}{$4\, ^1\! S_0$} &
\multicolumn{2}{c}{$5\, ^1\! S_0$} \\
\multicolumn{1}{c}{Initial} &
\multicolumn{1}{c}{$\lambda$ (\AA)} &
\multicolumn{1}{c}{$A$ (s$^{-1}$)} &
\multicolumn{1}{c}{$\lambda$ (\AA)} &
\multicolumn{1}{c}{$A$ (s$^{-1}$)} &
\multicolumn{1}{c}{$\lambda$ (\AA)} &
\multicolumn{1}{c}{$A$ (s$^{-1}$)} &
\multicolumn{1}{c}{$\lambda$ (\AA)} &
\multicolumn{1}{c}{$A$ (s$^{-1}$)} &
\multicolumn{1}{c}{$\lambda$ (\AA)} &
\multicolumn{1}{c}{$A$ (s$^{-1}$)}}
\startdata
\multicolumn{11}{c}{He-like carbon} \\
$2\, ^3\! P_1$&  40.751  &2.696[07]&      &  &  &       &	  &	     &         &	  \\
$3\, ^3\! P_1$&  35.086  &7.831[06]& 252.41   &4.524[05] &	  &	       &	  &	     &         &	  \\
$4\, ^3\! P_1$&  33.477  &3.289[06]& 187.57   &1.985[05] &  729.09& 6.230[04]  &	  &	     &         &	  \\
$5\, ^3\! P_1$&  32.786  &1.699[06]& 167.75   &1.035[05] &  499.63& 3.397[04]  & 1584.5   &1.517[04] &         &	  \\
$6\, ^3\! P_1$&  32.423  &1.086[06]& 158.68   &6.608[04] &  426.92& 2.187[04]  & 1028.8   &1.014[04] & 2927.8 & 5.502[03]\\[0.2pc]
\multicolumn{11}{c}{He-like nitrogen} \\
$2\, ^3\! P_1$&  29.095  &1.346[08]&	      & 	 &	  &	       &	  &	     &         &	  \\
$3\, ^3\! P_1$&  24.970  &3.984[07]&  176.50  &2.357[06] &	  &	       &	  &	     &         &	  \\
$4\, ^3\! P_1$&  23.802  &1.684[07]&  131.04  &1.041[06] &  508.76&  3.279[05] &	  &	     &         &	  \\
$5\, ^3\! P_1$&  23.301  &8.669[06]&  117.15  &5.415[05] &  348.40&  1.783[05] &  1104.6  &7.979[04] &         &	  \\
$6\, ^3\! P_1$&  23.038  &5.292[06]&  110.80  &3.312[05] &  297.61&  1.100[05] &  716.78  &5.108[04] &  2039.8 & 2.775[04]\\[0.2pc]
\multicolumn{11}{c}{He-like oxygen} \\
$2\, ^3\! P_1$&  21.810  &5.352[08]&	      & 	 &	  &	       &	  &	     &         &	  \\
$3\, ^3\! P_1$&  18.674  &1.605[08]&  130.28  &9.667[06] &	  &	       &	  &	     &         &	  \\
$4\, ^3\! P_1$&  17.788  &6.811[07]&  96.676  &4.290[06] &  374.99&  1.354[06] &	  &	     &         &	  \\
$5\, ^3\! P_1$&  17.407  &3.506[07]&  86.409  &2.231[06] &  256.69&  7.360[05] &  813.60  &3.298[05] &         &	  \\
$6\, ^3\! P_1$&  17.208  &2.093[07]&  81.710  &1.337[06] &  219.24&  4.451[05] &  527.78  &2.068[05] &  1501.8 & 1.124[05]\\[0.2pc]
\multicolumn{11}{c}{He-like neon} \\
$2\, ^3\! P_1$&  13.555  &5.253[09]&	      & 	 &	  &	       &	  &	     &         &	  \\
$3\, ^3\! P_1$&  11.570  &1.604[09]&  79.265  &9.896[07] &	  &	       &	  &	     &         &	  \\
$4\, ^3\! P_1$&  11.010  &6.840[08]&  58.785  &4.421[07] &  227.74&  1.399[07] &	  &	     &         &	  \\
$5\, ^3\! P_1$&  10.769  &3.523[08]&  52.528  &2.303[07] &  155.83&  7.608[06] &  493.68  &3.414[06] &         &	  \\
$6\, ^3\! P_1$&  10.644  &2.067[08]&  49.664  &1.357[07] &  133.06&  4.526[06] &  320.14  &2.105[06] &  910.80 & 1.145[06]\\[0.2pc]
\multicolumn{11}{c}{He-like silicon} \\
$2\, ^3\! P_1$&  6.6878  &1.550[11]&	      & 	 &	  &	       &	  &	     &         &	  \\
$3\, ^3\! P_1$&  5.6884  &4.798[10]&  38.172  &3.047[09] &	  &	       &	  &	     &         &	  \\
$4\, ^3\! P_1$&  5.4073  &2.054[10]&  28.299  &1.369[09] &  109.49&  4.337[08] &	  &	     &         &	  \\
$5\, ^3\! P_1$&  5.2867  &1.059[10]&  25.282  &7.145[08] &  74.902&  2.364[08] &   237.15 &1.061[08] &         &	  \\
$6\, ^3\! P_1$&  5.2236  &6.161[09]&  23.900  &4.179[08] &  63.950&  1.396[08] &   153.77 &6.498[07] &  437.31 & 3.536[07]\\[0.2pc]
\multicolumn{11}{c}{He-like argon} \\
$2\, ^3\! P_1$&  3.9685  &1.782[12]&	   &	      &        &	    &	       &	  &	    &	       \\
$3\, ^3\! P_1$&  3.3691  &5.486[11]&  22.341  &3.548[10] &        &	    &	       &	  &	    &	       \\
$4\, ^3\! P_1$&  3.2007  &2.343[11]&  16.563  &1.593[10] &  64.040&  5.046[09] &	       &	  &	    &	       \\
$5\, ^3\! P_1$&  3.1285  &1.206[11]&  14.796  &8.305[09] &  43.814&  2.748[09] &  138.68  &1.234[09] &	    &	       \\
$6\, ^3\! P_1$&  3.0907  &7.001[10]&  13.987  &4.848[09] &  37.408&  1.620[09] &  89.929  &7.542[08] &  255.70 & 1.620[09]\\
\enddata

\end{deluxetable}

\begin{deluxetable}{lrrrrrrrr}
\tablecaption{Wavelengths $\lambda$ (\AA) and radiative
transition rates $A$ (s$^{-1}$) for singlet-triplet transitions
in He-like ions. Numbers in brackets represent powers of 10.\label{xsp}}
\tablehead{
\multicolumn{1}{c}{Final} &
\multicolumn{2}{c}{$2\, ^3\! P_1$} &
\multicolumn{2}{c}{$3\, ^3\! P_1$} &
\multicolumn{2}{c}{$4\, ^3\! P_1$} &
\multicolumn{2}{c}{$5\, ^3\! P_1$}\\
\multicolumn{1}{c}{Initial} &
\multicolumn{1}{c}{$\lambda$ (\AA)} &
\multicolumn{1}{c}{$A$ (s$^{-1}$)} &
\multicolumn{1}{c}{$\lambda$ (\AA)} &
\multicolumn{1}{c}{$A$ (s$^{-1}$)} &
\multicolumn{1}{c}{$\lambda$ (\AA)} &
\multicolumn{1}{c}{$A$ (s$^{-1}$)} &
\multicolumn{1}{c}{$\lambda$ (\AA)} &
\multicolumn{1}{c}{$A$ (s$^{-1}$)}}
\startdata
\multicolumn{9}{c}{He-like carbon} \\
$3\, ^1\! S_0$&  252.51 & 1.209[05] &	     &  	 &	    &		&	  &	    \\
$4\, ^1\! S_0$&  187.60 & 4.961[04] & 730.80 & 3.665[04] &	    &		&	  &	    \\
$5\, ^1\! S_0$&  167.76 & 2.484[04] & 500.31 & 1.803[04] & 1589.25  &1.313[04]  &	  &	    \\
$6\, ^1\! S_0$&  158.68 & 1.418[04] & 427.37 & 1.010[04] & 1030.55  &7.273[03]  & 2937.4  &5.558[03]\\
\multicolumn{9}{c}{He-like nitrogen} \\
$3\, ^1\! S_0$&  176.14 & 6.094[05] &	     &  	 &	    &		&	  &	    \\
$4\, ^1\! S_0$&  130.85 & 2.502[05] & 508.97 & 1.898[05] &	    &		&	  &	    \\
$5\, ^1\! S_0$&  117.00 & 1.253[05] & 348.48 & 9.349[04] &  1105.9  &6.858[04]  &	  &	    \\
$6\, ^1\! S_0$&  110.66 & 7.156[04] & 297.66 & 5.239[04] &  717.26  &3.803[04]  & 2042.9  &2.900[04]\\
\multicolumn{9}{c}{He-like oxygen} \\
$3\, ^1\! S_0$&  129.87 & 2.446[06] &	     &  	 &	    &		&	  &	    \\
$4\, ^1\! S_0$&  96.463 & 1.005[06] & 374.80 & 7.768[05] &	    &		&	  &	    \\
$5\, ^1\! S_0$&  86.240 & 5.033[05] & 256.61 & 3.828[05] &  813.84  &2.823[05]  &	  &	    \\
$6\, ^1\! S_0$&  81.558 & 2.874[05] & 219.17 & 2.145[05] &  527.87  &1.566[05]  & 1502.8  &1.194[05]\\
\multicolumn{9}{c}{He-like neon} \\
$3\, ^1\! S_0$&  78.964 & 2.442[07] &	     &  	 &	    &		&	  &	    \\
$4\, ^1\! S_0$&  58.632 & 1.003[07] & 227.482 & 7.953[06] &	    &		&	  &	    \\
$5\, ^1\! S_0$&  52.407 & 5.025[06] & 155.721 & 3.920[06] &  493.51  &2.911[06]  &	  &	    \\
$6\, ^1\! S_0$&  49.556 & 2.870[06] & 132.991 & 2.198[06] &  320.08  &1.616[06]  & 910.78  &1.234[06]\\
\multicolumn{9}{c}{He-like silicon} \\
$3\, ^1\! S_0$&  38.058 & 7.402[08] &	     &  	 &	    &		&	  &	    \\
$4\, ^1\! S_0$&  28.241 & 3.037[08] & 109.41 & 2.461[08] &	    &		&	  &	    \\
$5\, ^1\! S_0$&  25.236 & 1.521[08] & 74.869 & 1.213[08] &  237.13  &9.057[07]  &	  &	    \\
$6\, ^1\! S_0$&  23.859 & 8.682[07] & 63.926 & 6.800[07] &  153.76  &5.028[07]  & 437.40  &3.845[07]\\
\multicolumn{9}{c}{He-like argon} \\
$3\, ^1\! S_0$&  22.304 & 8.679[09] &	     &  	 &	    &		&	  &	    \\
$4\, ^1\! S_0$&  16.543 & 3.557[09] & 64.050 & 2.881[09] &	    &		&	  &	    \\
$5\, ^1\! S_0$&  14.781 & 1.780[09] & 43.818 & 1.420[09] &  138.77  &1.058[09]  &	  &	    \\
$6\, ^1\! S_0$&  13.973 & 1.016[09] & 37.409 & 7.957[08] &  89.963  &5.875[08]  & 255.92  &4.488[08]
\enddata
\end{deluxetable}


\end{document}